%% file: p.tex
\documentclass[12pt]{article}
\usepackage{graphicx}

\newcommand\pubnumber{Article 29 in eConf C1304143}
\newcommand\pubdate{\today}

\def\iap{Institut d'Astrophysique de Paris, 98Bis Bd Arago, 75014 Paris, France}

\def\Title#1{\begin{center} {\Large #1 } \end{center}}
\def\Author#1{\begin{center}{ \sc #1} \end{center}}
\def\Address#1{\begin{center}{ \it #1} \end{center}}

\def\spose#1{\hbox to 0pt{#1\hss}}
\def\lesssim{\mathrel{\spose{\lower 3pt\hbox{$\mathchar"218$}}
             \raise 2.0pt\hbox{$\mathchar"13C$}}}

\newcommand\pubblock{\rightline{\begin{tabular}{l} \pubnumber\\
         \pubdate  \end{tabular}}}
\newenvironment{Abstract}{\begin{quotation}  }{\end{quotation}}
\newenvironment{Presented}{\begin{quotation} \begin{center} 
             PRESENTED AT\end{center}\bigskip 
      \begin{center}\begin{large}}{\end{large}\end{center} \end{quotation}}
\def\Acknowledgements{\bigskip  \bigskip \begin{center} \begin{large}
             \bf ACKNOWLEDGEMENTS \end{large}\end{center}}

\input econfmacros.tex

\begin{document}
\begin{titlepage}
\pubblock

\vfill
\Title{The prompt-early afterglow connection in GRBs}
\vfill
\Author{Robert Mochkovitch \\ Fr\'ed\'eric Daigne}
\Address{\iap}
\Author{Romain Hasco\"et}
\Address{Physics Department and Columbia Astrophysics Laboratory, Columbia University, 538 West
120th Street New York, NY 10027}
\vfill
\begin{Abstract}
We study the observed correlations between the duration and luminosity of the early afterglow plateau 
and the isotropic gamma-ray energy release during the prompt phase. We discuss these correlations in the
context of two scenarios for the origin of the plateaus. In the first one the afterglow is made by the
forward shock and the plateau results from variations of the microphysics parameters while in the second
one the early afterglow is made by a long-lived reverse shock propagating in a low $\Gamma$ tail of the
ejecta. 

\end{Abstract}
\vfill
\begin{Presented}
``Huntsville in Nashville"
Gamma-Ray Burst Symposium\\
Nashville, USA,  April 14--18, 2013
\end{Presented}
\vfill
\end{titlepage}
\def\thefootnote{\fnsymbol{footnote}}
\setcounter{footnote}{0}

\section{Introduction}
Before the launch of the {\it Swift} satellite in 2004~\cite{Gehrels}, the afterglow was believed to be the best understood part
of the GRB phenomenon, being explained by the energy dissipated in the forward shock formed by the jet impacting
the burst environment. However, the many surprises of the early X-ray afterglow revealed by {\it Swift} -- initial steep decay,
plateau phase, flares~\cite{Nousek} -- have considerably complicated the picture. Several mechanisms have been proposed to explain the plateau 
but none appears fully convincing. 

The recent discovery of correlations linking prompt and early afterglow quantities~\cite{Dainotti,Margutti} such as
$E_{\gamma,{\rm iso}}$, $t_{\rm p}$, the duration of the plateau and $L_{\rm p}$, the plateau luminosity, represents new constraints 
to be satisfied by models. After a critical discussion of the case where the plateau is made by late energy injection into the forward shock  
we consider two alternative scenarios and check if they can agree with the observed correlations.

\section{Making a plateau with late energy injection}
Continuous energy injection into the forward shock remains a commonly invoked cause of plateau formation. For the most extended plateaus
it however imposes to inject several hundreds times more energy than was initially present to power the prompt phase. 
This huge amount of energy leads to an ``efficiency crisis'' for the prompt mechanism,
whatever it is. Let $E_0$ and $E=k\,E_0$ be respectively the values of the energy available during the prompt phase and after injection.
Then, the measured gamma-ray efficiency is
\begin{equation}
f_{\gamma,{\rm mes}}={E_\gamma\over E_\gamma+E}
\end{equation}
because the energy in the forward shock is estimated from multiwavelength fits of the afterglow after typically one day (i.e. after 
energy injection). However the true efficiency
\begin{equation}
f_{\gamma,{\rm true}}={E_\gamma\over E_\gamma+E_0}={1\over 1+{1\over k}\left({1\over f_{\gamma,{\rm mes}}}-1\right)}
\end{equation} 
can be much larger. With for example $f_{\gamma,{\rm mes}}=0.1$, the true efficiency is $f_{\gamma,{\rm true}}=0.53$ for $k=10$ and $0.92$ for $k=100$.
These values of $f_{\gamma,{\rm true}}$ seem unreachable for any of the proposed prompt mechanisms: the efficiency of internal shocks can barely reach 10\% while
that of comptonized photosphere or reconnection models is more uncertain but probably does not exceed 30\%. 
\section{Making a plateau without energy injection}
\subsection{Within the standard forward shock scenario}
Without energy injection the standard forward shock scenario can successfully account for the afterglow evolution after 
about one day but fails to reproduce the plateau phase. A backwards extrapolation of the late afterglow flux lies above the plateau, which might
therefore be interpreted as the indication that some radiation is ``missing''. This can be the case if the radiative efficiency 
of the forward shock during the early afterglow is smaller than predicted by the standard model. The most obvious way to reduce the efficiency 
is to relax the assumption that the microphysics parameters stay constant throughout the whole afterglow evolution~\cite{Granot,Ioka}. 
For both a uniform and a wind external medium the afterglow X-ray flux behaves as 
\begin{equation}
F_{\rm X}\propto E^{p+2\over 4}\epsilon_e^{p-1}\epsilon_B^{p-2\over 4}t^{-{3p-2\over 4}}
\end{equation}    
where $E$ is the burst isotropic energy, $\epsilon_e$ and $\epsilon_B$ the microphysics parameters and $p$ the power-law index of the
accelerated electron spectrum. With $2<p<3$ the dependence on $\epsilon_B$ is weak so that in practice 
only playing with $\epsilon_e$ can really affect the flux evolution. A priori $\epsilon_e$ can be a function of the shock Lorentz factor, the density of
the external medium (if it is a stellar wind) or both. 

The stellar wind case is especially interesting if one assumes that, below a critical
density $n_0$, $\epsilon_e$ is constant while $\epsilon_e\propto n^{-\nu}$ (with $\nu>0$) for $n>n_0$. 
Since the density seen by the forward shock is given by $n(t)\sim 6.3\,10^4 A_*^2 E_{53}^{-1} t^{-1}$ cm$^{-3}$ (where $A_*$ is the wind parameter; see~\cite{pk})
the transition at $n_0$, which corresponds to the end of the plateau, takes place at
\begin{equation}
t_{\rm p}\approx 6.6\,10^4 A_*^2 n_0^{-1} E_{53}^{-1}\approx 6.6\,10^4 A_*^2 n_0^{-1} f_{\gamma} E_{\gamma,53}^{-1}\ \ {\rm s}
\end{equation}  
where $f_{\gamma}$ is the gamma-ray efficiency of the prompt phase and $E_{\gamma,53}$ is the isotropic gamma-ray energy release. 
If the product $A_*^2 n_0^{-1} f_{\gamma}$ does not vary much from burst to burst (and stays close to about $0.1$)
Eq.(4) is not too different from the observed [$t_{\rm p},E_{\gamma,{\rm iso}}]$ relation.

A flat plateau is obtained for $\nu={3p-2\over 4(p-1)}=\nu_0=0.92$ if $p=2.5$
while for $\nu<\nu_0$ (resp. $\nu>\nu_0$) the plateau flux is decreasing (resp. rising) with time. 
Since $n(t)\propto t^{-1}$,
a flat plateau extending over two decades in time (as in GRB 060729) requires an increases of $\epsilon_e$ by a factor of about $100$ from the
beginning to the end of the plateau. 
\subsection{With a long-lived reverse shock}
We now suppose that the ejecta emitted by the central engine is made of a ``head'' 
with material at high Lorentz factors ($\Gamma \sim 10^2$ - $10^3$), followed by a ``tail''
where the Lorentz factor decreases to much smaller values, possibly close to unity. The head 
is responsible for the prompt emission while the reverse shock propagating through the tail 
makes the afterglow~\cite{Genet,Uhm}.

We adopt for the head a constant energy injection rate ${\dot E}_{\rm H}$ 
for a duration of 10 s. We do not specify the distribution of the Lorentz factor 
and simply consider its average value, supposed to be $\bar \Gamma=400$. 
The tail that follows lasts for $100$ s but this value is not critical
as long as it does not exceed the duration of the early steep decay phase observed at the
beginning of most X-ray light curves. 
We start with a simple case where the distribution of energy in the tail 
${dE\over d{\rm Log}\Gamma}$ is constant from $\Gamma=400$ to $\Gamma=1$. This can be obtained by adopting a constant 
energy injection rate ${\dot E}_{\rm T}$ and 
a Lorentz factor of the form
\begin{equation} 
\Gamma_{\rm T}(s)=400^{\left({1.1-s/100}\right)}\ ,
\end{equation}
from $s=10$ to $110$ light.seconds, the distance $s$ being counted from the front to the back of the flow.

Using the methods described in~\cite{Genet} we have obtained the power $P_{\rm diss}(t)$ dissipated by the reverse shock  
(as a function of arrival time to the observer) for ${\dot E}_{\rm H}=10{\dot E}_{\rm T}$ (so that equal amounts
of energy are injected in the head and tail, and two possibilities for the burst environment: ({\it i}) a uniform medium 
with $n=1$ cm$^{-3}$  
or ({\it ii}) a stellar wind with a wind parameter $A_*=1$.

Going from the dissipated power to actual light curves depends on assumptions
that have to be made for the microphysics parameters but the general shape 
of the early X-ray afterglow light curves remains globally similar to the evolution of $P_{\rm diss}(t)$  
so that some conclusions can already be reached without having to consider the uncertain post-shock microphysics.    

If energy is evenly distributed in the tail (constant ${dE\over d{\rm Log}\Gamma}$) 
a short plateau lasting about $1000 s$ is observed for a uniform external medium while for a stellar wind
there is no plateau. In both cases, 
the dissipated power approximately decays as
$t^{-1}$ after about 1000 s. It is larger by a factor 3 - 5 for a uniform medium due to a larger contrast $\kappa$ of the Lorentz 
factors at the reverse shock compared to the wind case ($\kappa\simeq 2$ and $2^{1/2}$ respectively, see~\cite{Genet}),
leading to a higher efficiency.  

We now vary the energy deposition in the tail, concentrating more power at some value of the Lorentz factor.
We have for example considered a simple model where
\begin{equation}
{\dot E}_{\rm T}(\Gamma)= \left\lbrace\begin{array}{cl}
{\dot E}_*\left({\Gamma\over \Gamma_*}\right)^{-q}\ \ \ {\rm for}\ \ \Gamma>\Gamma_*\\
{\dot E}_*\left({\Gamma\over \Gamma_*}\right)^{q^{\,\prime}}\ \ \ {\rm for}\ \ \Gamma<\Gamma_*\\
\end{array}\right.
\end{equation}
the value of ${\dot E}_*$ being fixed by the total energy injected in the tail. 
When energy deposition is more concentrated (increasing $q$) a plateau progressively forms and becomes flatter.
The value of $\Gamma_*$ in Eq.(6) fixes the duration $t_{\rm p}$ of the plateau as it approximately corresponds to the time when the reverse shock
reaches $s_*$ where $\Gamma_{\rm T}(s_*)=\Gamma_*$. The $q$ parameter controls the flatness of the plateau while $q^{\,\prime}$ controls the decay 
index after the plateau. As long as $E_{\rm T} \lesssim E_{\rm H}$, the deceleration of the head can be described as if it does not receive
any supply of energy from the tail, which gives 
\begin{equation}
t_{\rm p}\sim \left\lbrace
\begin{array}{cl}
& 9\,10^4\,E_{{\rm H},53}^{1/3}\,n^{-1/3}\,\Gamma_{*,1}^{-8/3}\ \ \ {\rm s}\ \ \ {\rm (uniform\ medium)}\\
& 3\,10^4\,E_{{\rm H},53}\,A_*^{-1}\,\Gamma_{*,1}^{\,-4}\ \ \ {\rm s}\ \ \ {\rm (stellar\ wind)}\\
\end{array}\right.
\end{equation}
Then, an approximate analytical solution for the power dissipated in the reverse shock can be obtained following the method 
described in~\cite{Genet}
\begin{equation}
P_{\rm diss}(t)={E_{\rm T}\over t_{\rm p}\,\varphi_{qq^{\,\prime}}} F(\gamma)\,\left({t\over t_{\rm p}}\right)^{\pm \,q\gamma-1}\ ,
\end{equation}
where $\varphi_{qq^{\,\prime}}={1\over q}+{1\over q^{\,\prime}}$ and $F(\gamma)={\gamma\over 2}\left[1-(1-2\gamma)^{1/2}\right]^{2}$
with $\gamma=3/8$ (resp. $1/4$) for a uniform medium (resp. a stellar wind).
The decay indices before and after the break at the end of the plateau are
\begin{equation}
\begin{array}{cl}
& \alpha_1=\gamma q-1\\
& \alpha_2=-\gamma q^{\,\prime}-1
\end{array}
\end{equation}
so that a flat plateau   is expected for $q=1/\gamma$ (i.e. $q=8/3$ and $4$ in the uniform medium and
wind cases ({\it i}) and ({\it ii}) respectively). 
For a typical decay index $\alpha_2=-1.5$ after the plateau
we get the condition $q^{\,\prime}=1/2\gamma$ (i.e. $q^{\,\prime}=4/3$
and $2$ for cases ({\it i}) and ({\it ii})). 
\section{Building a sequence of models}
\subsection{Forward shock scenario}
It has been shown in Sect.3.1 that a transition in the behavior of $\epsilon_e$ (from rising to constant) at a fixed density $n_0$ marks the end of the
plateau at a time $t_{\rm p}$ given by Eq.(4). The X-ray luminosity $L_{\rm p}$ at $t=t_{\rm p}$ then reads 
\begin{equation}
L_{\rm p}\propto E^{p+2\over 4}t_{\rm p}^{-{3p-2\over 4}}\propto t_{\rm p}^{-p}\propto E_{\gamma,{\rm iso}}^p
\end{equation}
as long as the efficiency and the microphysics parameters do not strongly vary from burst to burst. We have obtained a sequence of afterglow light curves
for $\epsilon_e=0.1$, $\epsilon_B=0.01$, $n_0=1.7$ cm$^{-3}$, $f_{\gamma}=0.1$, $A_*=1$ and $p=2.5$ and different values of the isotropic
gamma-ray energy release $E_{\gamma,{\rm iso}}$ from which we have plotted the three relations [$L_{\rm p},E_{\gamma,{\rm iso}}$], [$L_{\rm p},t_{\rm p}$] 
and [$L_{\rm p}/E_{\gamma,{\rm iso}},t_{\rm p}$] shown in Fig.1.  
\subsection{Reverse shock scenario}
Using Eq.(7) it is possible to link the duration of the plateau to the gamma-ray energy release $E_{\gamma,{\rm iso}}$, for a given dependence of
$\Gamma_*$ on the burst energy. Adopting $\Gamma_*\propto E^{1/2}$ as suggested in~\cite{Ghirlanda} (see however~\cite{hasc})
leads to
\begin{equation}
t_{\rm p}\propto E^{-1}\propto E_{\gamma,{\rm iso}}^{-1}
\end{equation}
for both a uniform medium and a stellar wind (if the gamma-ray efficiency does not depend on $E$). 
Together with Eq.(8)
this fixes the dissipated power during the plateau phase
\begin{equation}
P_{\rm diss}\propto t_{\rm p}^{-2}\propto E_{\gamma,{\rm iso}}^2
\end{equation}
To now compute a sequence of X-ray light curves from the dissipated power we need the microphysics parameters $\epsilon_e$ and 
$\epsilon_B$ for which we adopt the fiducial values $\epsilon_e=0.1$ and
$\epsilon_B=0.01$. From this sequence of light curves, we obtain $t_{\rm p}$ and $L_{\rm p}$ as a 
function of $E_{\gamma,{\rm iso}}$ and again plot the results in the three diagrams of Fig.1. 
\begin{figure}[htb]
\centering
\begin{tabular}{ccc}
\includegraphics[height=1.86in]{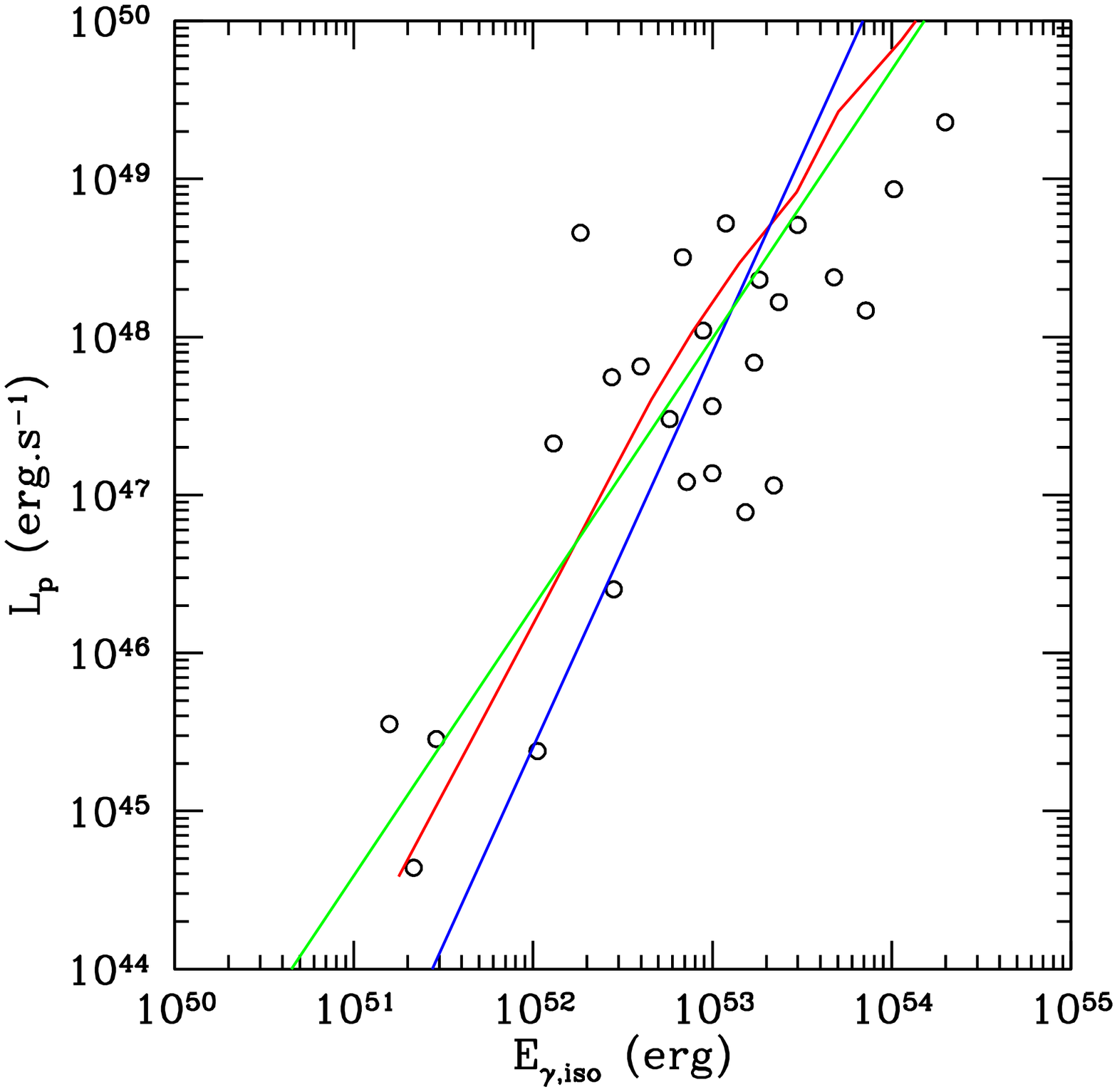}&
\includegraphics[height=1.86in]{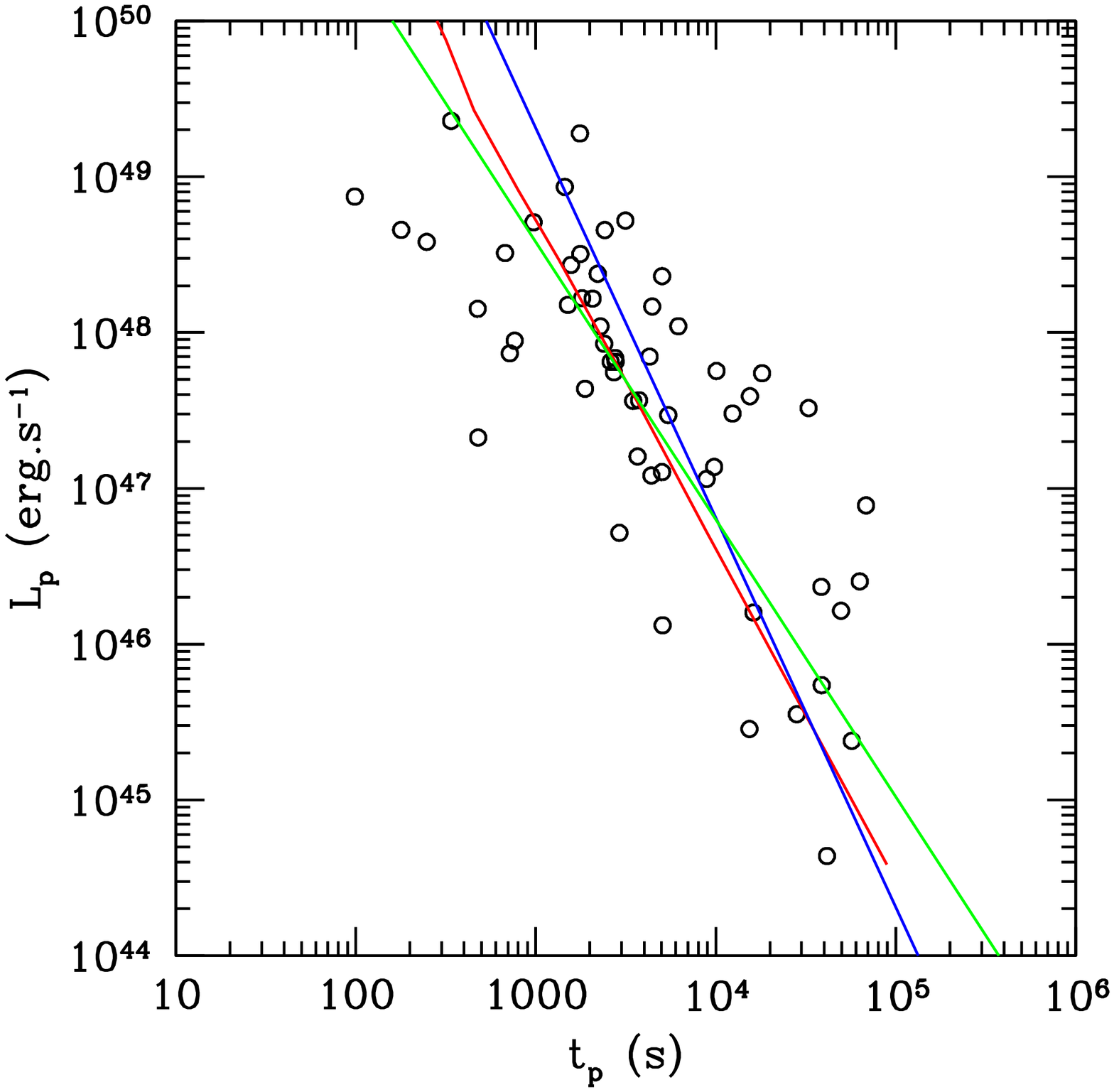}&
\includegraphics[height=1.86in]{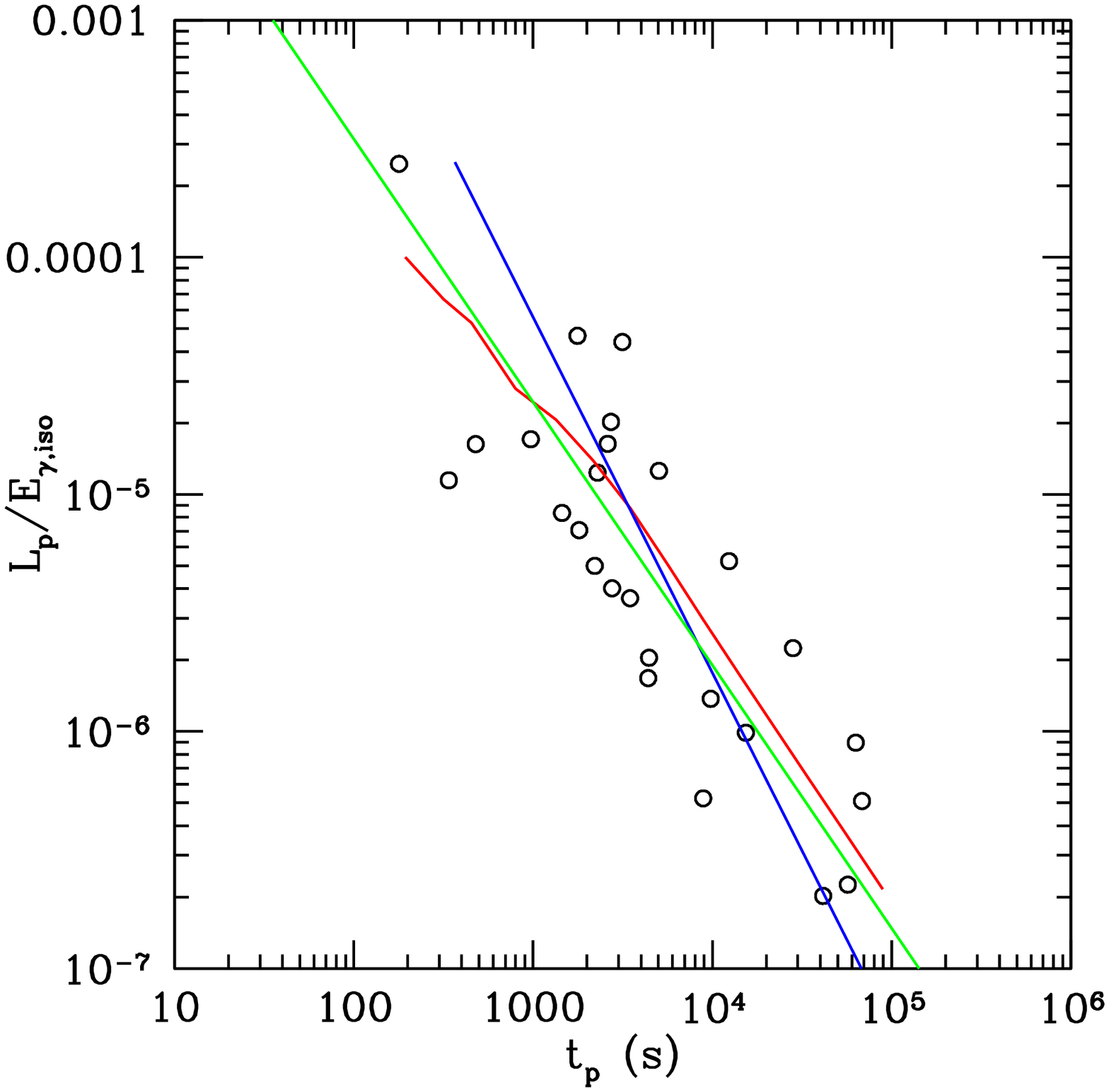}
\end{tabular}
\caption{Model [$L_{\rm p},E_{\gamma,{\rm iso}}$], [$L_{\rm p},t_{\rm p}$], and [$L_{\rm p}/E_{\gamma,{\rm iso}},t_{\rm p}$] correlations
compared to the data collected by~\cite{Margutti}. The green line is the best power-law fit of the data while the blue and red ones
respectively correspond to the forward and reverse shock scenarios. }
\end{figure}

\section{Discussion and conclusion}
The two scenarios we have considered to account for the origin the X-ray plateau 
seem able to explain the main features of the prompt-afterglow connection.  When compared to data 
the [$L_{\rm p},E_{\gamma,{\rm iso}}$] and [$L_{\rm p},t_{\rm p}$] correlations however appear somewhat steeper,
especilly in the forward shock case. The forward shock scenario also imposes a wind external medium and
that $\epsilon_e$ first increases with decreasing wind density.
In the reverse shock case, the shock propagates in a low-$\Gamma$ tail of the ejecta with a peak of 
injected power at a Lorentz factor $\Gamma\sim 20$ - 30. This scenario supposes that the forward shock is 
(at least initially) radiatively inefficient, as suggested by~\cite{ml}. It may look more exotic but has an advantage of flexibility 
as various accidents in the early X-ray afterglow (such as steep breaks, bumps,...) can be simply produced 
by playing with the distribution of injected power in the tail (see~\cite{Uhm2}).

\Acknowledgements
It is a pleasure to thank Raffaella Margutti who kindly sent us her data on the prompt-afterglow correlations. 
R.H. acknowledges the support by NSF grant AST-1008334.

\end{document}

%% file: econfmacros.tex
\def\beq{\begin{equation}}
\def\eeq#1{\label{#1}\end{equation}}
\def\eeqn{\end{equation}}

\def\beqa{\begin{eqnarray}}
\def\eeqa#1{\label{#1}\end{eqnarray}}
\def\eeqan{\end{eqnarray}}

\let\bar=\overbar

\def\Dslash{\not{\hbox{\kern-4pt $D$}}}
\def\dslash{\not{\hbox{\kern-2pt $\del$}}}

\def\msb{{\bar{\ssstyle M \kern -1pt S}}}

